\documentstyle[preprint,aps,prl,epsfig,multicol,float]{revtex}

\begin{document}

\title{Superconductivity in Magnetically Ordered CeTe$_{1.82}$}

\author{M. H. Jung$^{1}$, A. Alsmadi$^{2}$, H. C. Kim$^{3}$,  Yunkyu Bang$^{4}$, 
K. H. Ahn$^{4}$, K. Umeo$^{5}$, A. H. Lacerda$^{1}$, H. Nakotte$^{2}$, 
H. C. Ri$^{3}$, and T. Takabatake $^{5}$}

\address{$^{1}$ National High Magnetic Field Laboratory, Los Alamos National Laboratory,
Los Alamos NM 87545, USA \\
$^{2}$ Department of Physics, New Mexico State University, Las Cruces NM 88003,
USA \\
$^{3}$ Material Science Laboratory, Korea Basic Science Institute, Taejeon
305-333, Korea \\
$^{4}$ Theoretical Division, Los Alamos National Laboratory, Los Alamos NM
87545, USA \\
$^{5}$ Department of Quantum Matter, ADSM, Hiroshima University,
Higashi-Hiroshima 739-8530, Japan\\}


\maketitle

\begin{abstract}

We report the discovery of pressure-induced superconductivity in a
semimetallic magnetic material CeTe$_{1.82}$. The superconducting transition 
temperature $T_{SC}$ = 2.7 K (well below the magnetic ordering temperatures) 
under pressure ($>$ 2 kbar) is remarkably high, considering the relatively low carrier density 
due to a charge-density-wave transition associated with lattice modulation.
The coexisting magnetic structure of a mixed ferromagnetism 
and antiferromagnetism can provide a clue for this high $T_{SC}$. 
We discuss a theoretical model for its possible pairing symmetry and
pairing mechanism.

\end{abstract}

\noindent 


The $f$-electron compounds never cease to surprise us with interesting ground states. 
Due to the local nature of $f$-obital wave function, these compounds display various 
many-body collective state(s) at low temperature -- often competition/coexistance among them.
Typically, the $f$-electron systems show a heavily renormalized quasiparticle (called 
heavy fermion) below a certain crossover temperature, and then many of them at lower 
temperatures show magnetism and/or superconductivity (SC) such as in CeCu$_2$Si$_2$, 
CeIn$_3$, CeRhIn$_5$, UNi$_2$Al$_3$, UGe$_2$, etc. \cite{HF_SC,QCP}. 
CeTe$_{1.82}$ studied here certainly belongs to this class of materials.
However, due to Te and the crystal structure this compound also shares some of the feature 
of the layered transition-metal dichalcogenides and NbSe$_3$, which undergo a 
superconducting transition at low temperature $\sim$ 1 K with the charge-density-wave (CDW) 
ordering at far higher temperature $\sim$ 1000 K \cite{CDW_SC}. The precise role of CDW 
with respect to SC is unclear so far and their coherent properties now constitute 
a separate interesting branch of correlated electron systems \cite{Neto}. 
The compound CeTe$_{1.82}$ studied in this Letter shows all these collective 
states: CDW, magnetism, and SC in consecutively lowering temperature.

Here we report the first observation of pressure-induced SC in a
semimetallic magnetic material CeTe$_{1.82}$ with a relatively low density of
states (DOS) \cite{DOS}. At ambient pressure, CeTe$_{2-\delta} (0.13 \leq \delta \leq 
0.18)$ displays various collective ground states and exhibits highly anisotropic 
transport and magnetic properties. CeTe$_{2-\delta}$ crystallizes in layered 
tetragonal Cu$_2$Sb-type structure, where a metallic Te sheet is sandwiched by 
semiconducting CeTe double layers and is stacked along the $c$ axis \cite{Jung_xtal}. 
Because of this layered crystal structure and the Te vacancy, a CDW state is stabilized 
even at far above the room temperature. The presence of CDW gap ($T_{CDW} \sim$ 1000 K) accompanying 
with a lattice distortion in the Te sheet is verified by electron-tunneling 
spectroscopy measurements \cite{Jung_tunneling}. At low temperatures, this compound 
undergoes two different magnetic orderings. The local magnetic moments of Ce ions 
develop a short-range ferromagnetic (SRF) ordering in the CeTe layer with a magnetoelastic 
origin below $T_{SRF} \sim$ 6 K. As temperature is further lowered, the SRF CeTe layers 
develop a long-range ferromagnetic (FM) order in the layers and simultaneously a 
long-range antiferromagnetic (AFM) order in the spin sequence of down-up-up-down along 
the $c$ axis below $T_{N} \sim$ 4.3 K (see Ref. \cite{Jung_phase} and Fig. 1). 
Because of the two-dimensional motion of the carriers confined within the Te sheet 
sandwiched by the ferromagnetically coupled CeTe layers, the strong anisotropy is observed 
in the electrical resistivity with a ratio of $\rho_{\parallel c}/\rho_{\perp c} \sim$ 150 
at 2 K and the isothermal magnetization with a ratio of $M_{\parallel c}/M_{\perp c} \sim$ 7 
at 2 kG \cite {Jung_xtal,Jung_phase}.

High-purity single crystals were grown with varying Te contents, i.e., 
$0.13 \leq \delta \leq 0.18$ in CeTe$_{2-\delta}$. Electron-probe microanalysis 
reveals the deficiency in the Te content, $\delta$, without any evidence of inhomogeneity 
to a resolution of 0.1\%. This $\delta$ value is often observed in other rare-earth 
dichalcogenides such as LaTe$_{1.9}$, SmTe$_{1.84}$, and DySe$_{1.85}$ \cite{delta_dichal}, 
where the chalcogen vacancy goes into the Te sheet and stabilizes the structural 
modulation. The in-plane resistivity measurements were made on
single-crystal platelets by the conventional a.c. four-terminal method as a
function of temperature, magnetic field, and pressure. The pressure cell is of
the piston-cylinder type constructed out of high-purity non-magnetic BeCu 
alloy suitable for the application of external magnetic fields. The pressure
was determined to $\pm$ 0.005 kbar from the electrical resistance of Manganin
sensor. Bulk magnetization measurements as a function of temperature were
performed by means of a SQUID magnetometer (Quantum Design, MPMS7) with similar
pressure cells, in which 1:1 mixture of Flurinert FC70 and FC77 was used for a
pressure transmitting medium.

Figure 2 displays a typical feature of $\rho(T)$ showing the superconducting 
transition in CeTe$_{1.82}$ at $P$ = 5 kbar, where $\rho(T)$ starts to 
drop drastically at $T_{SC}$ = 2.7 K. The application of 
magnetic field suppresses the resistivity drop, as expected for a 
superconducting transition. From the onset of the superconducting transition, 
we have determined the superconducting phase diagram shown in the inset of Fig. 2. 
It is rather unusual that the upper critical field $H_{c2}$ ($\sim$  5 kG at
$T \rightarrow 0$) is about an order smaller than $H_{c2}(0)$ ($\sim$
100 kG at 20 kbar) of CeRhIn$_5$ \cite{Hc2} having a similar $T_{SC}$ ($\sim$ 2.1 K). 
We consider the low DOS of CeTe$_{1.82}$ as the primary reason for such a small $H_{c2}$. 
Then, the relatively high $T_{SC}$ with a low DOS indicates a different pairing 
mechanism compared to CeRhIn$_5$ and other heavy fermion superconductors.

A more conclusive evidence for SC is a diamagnetic signal below $T_{SC}$, 
and thus we measured the magnetization $M(T)$ of CeTe$_{1.82}$ for different pressures. 
Because the superconducting transition occurs just below the magnetic transition and 
the SC coexists with the magnetism below $T_{SC}$, a diamagnetic signal 
associated with SC is quite small to be easily detected. Hence, we have first 
plotted the difference $\Delta M$ (= $M_{ZFC}-M_{FC}$) between the zero-field-cooled 
($M_{ZFC}$) and field-cooled data ($M_{FC}$) in the inset of Fig. 3, 
showing a clear deviation from the linear-temperature-dependent background 
($M_{BG}$). This background is defined as an extrapolation of the magnetic 
hysteresis of $M(T)$ below $T_N$.  
The main panel of Fig. 3 shows the magnetic susceptibility  
$4 \pi \Delta M_{SC} / H$ measured at 6 kbar, where $\Delta M_{SC} = \Delta M-M_{BG}$, 
assuming the density of CeTe$_{1.82}$ is about 10 g/cc. The diamagnetic component is 
observed just below 2.8 K, which coincides with $T_{SC}$ determined by $\rho(T)$ 
measurements. These results support the presence of bulk SC in CeTe$_{1.82}$.

In Fig. 4, we draw the phase diagram in temperature-pressure space summarizing
our measurements for CeTe$_{1.82}$. $\rho(T)$ and $M(T)$ at different pressures allow us to
identify the short-range ferromagnetic ordering temperature $T_{SRF}$ and the
long-range ferro/antiferro-magnetic ordering temperature $T_N$. The
superconducting transition temperature $T_{SC}$ is determined from $\rho(T)$.
The applied pressure slightly enhances both $T_{SRF}$ and $T_N$ over the whole
region of measured pressure. The SC suddenly appears in the narrow region below 2 kbar. 
The pressure-induced SC often occurs in heavy fermion metals
in the vicinity of AFM or FM quantum criticality (QC: $T_N$ or $T_C \rightarrow$ 0 K) 
\cite{QCP} and the normal state properties exhibit various deviations
from Fermi-liquid metal, so called, non-Fermi liquid (NFL) behavior \cite{NFL}. 
However, for CeTe$_{1.82}$ there is no magnetic QC in our phase diagram 
and the superconducting phase exists completely inside the magnetic phase. 
Also, the transport and magnetic properties show no NFL behavior.

We examine possible theoretical scenarios for the SC in CeTe$_{1.82}$,
focusing on pairing interactions and pairing symmetries.
If the SC is mediated by magnetic fluctuations, the phase diagram 
in Fig. 4 appears consistent with a FM-induced SC; 
an AFM-induced SC tends to appear near the boundary between magnetic
and non-magnetic phases \cite{ferro_SC}. 
In fact, considering the ferro/antiferro-magnetic ordering 
structure (Fig.1), in which the main conducting Te sheet is sandwiched 
between two FM CeTe double layers and this FM sandwich structure is 
alternating its polarization along the $c$ axis, it is quite plausible that the
carriers confined in the Te sheet interact by exchange of FM fluctuations 
and form superconducting pairs \cite{ferro_SC}. While the traditional idea for the 
FM-induced SC is a triplet and odd orbital pairing, recent theoretical studies suggest 
that singlet s-wave pairing is also possible inside a FM phase \cite{ferro_SC}.
Although it remains an important issue for further experiments to determine the
symmetry of the superconducting order parameter, the possibility of FM 
triplet pairing in CeTe$_{1.82}$ has a couple of problems because of its sensitivity to disorder.
The sample which exhibits SC has Te vacancy of $\sim 10\%$, and most of this vacancy is
believed to go into the Te sheets that are going to develop SC. Any triplet
odd orbital pairing hardly survives in this much disorder.
In addition, the pressure-induced SC is observed only for a single
crystal with $\delta$ = 0.18 in CeTe$_{2-\delta}$. We have measured in-pressure
resistivity of other single crystals with $\delta$ = 0.15 and 0.13, 
and found almost identical magnetic properties but no
superconducting transition. This indifference of SC to the
magnetic properties and the extreme sensitivity to the Te vacancy
suggests that the magnetism is unlikely to be a primary
source of SC pairing mechanism.

From the sensitive dependence of SC to $\delta$, one could 
speculate that the SC in CeTe$_{1.82}$ is most likely to be
associated with the crystal-lattice instability. The existence of CDW
instability driven by the Fermi surface nesting is realized with a small
periodic lattice distortion within the Te sheet, which is stabilized by a vacancy
order in the Te sheets \cite{Jung_tunneling}. 
Related to this, an interesting pairing mechanism has been proposed 
by Castro Neto \cite{Neto} in order to explain 
CDW-SC in transition-metal dichalcogenides (TMD).
In this theory, the SC pairing occurs with the Dirac 
fermions formed after a gapless CDW ordering, which couple with acoustic phonons
via piezoelectric coupling due to the inversion symmetry breaking by a six-fold
CDW order. The main difference between TMD compounds and CeTe$_{2-\delta}$ is
that CeTe$_{2-\delta}$ has $f$-orbital moments from Ce ions and these moments
develop  magnetic orderings at low temperature (6 K and 4.3 K). There are
experiments indicating possible interplay between CDW and magnetic order 
in CeTe$_{2-\delta}$, but
the details are still unclear \cite{Jung_tunneling}. Also the piezoelectric 
coupling is, in general, unlikely in metals\cite{Varma_comment}. 
Therefore, the application
of the pairing theory for TMD to our case is not
straightforward. However, on a general ground the CDW ordering and accompanying
lattice modulation should create a new optical phonon mode, 
which then couples to electrons in the Te layers.
Hence, we speculate that the primary pairing interaction is mediated by phonons forming a
s-wave singlet SC.

As for the role of magnetism for SC, we can think of two
effects: (1) In addition to the phonon pairing potential, the FM 
fluctuations in the FM phase can contribute to a s-wave singlet pairing
\cite{ferro_SC}; (2) The tunneling data from Ref. \cite{Jung_tunneling} shows
that the zero-bias conductance increases below $T_{SRF}$, indicating that DOS
increases due to magnetic ordering. 
Finally as for the role of pressure, we think that 
the $c$-axis lattice distance is crucial to determine the actual
$T_{SC}$ as in TMD. Figure 4 shows that the SC, coexisting with
magnetism, abruptly appears at $T_{SC}$ = 2.7 K with as low  pressure as 2
kbar. As indicated by the dotted line, we cannot rule out that the superconducting phase
appears sharply below 2 kbar.  Then the pressure might decrease the $c$ lattice 
constant and this will increase the interlayer coupling to increase $T_{SC}$ 
as in other layered superconductors.

To conclude, we report the superconducting transition at $T_{SC}$ = 2.7 K in
CeTe$_{1.82}$ under pressure ($P >$ 2 kbar). CeTe$_{2-\delta}$ displays various 
collective states at different temperatures; CDW ($\sim$ 1000 K), SRF ($\sim$ 6 K), 
$c$-axis AFM ($\sim$ 4.3 K), and finally SC transition for $\delta$ = 0.18. 
Combining available data and phase diagram, 
we conclude that the primary pairing mechanism is a 
phonon-mediated s-wave SC, enhanced by the FM fluctuations inside the
magnetic ordering phase and the increased DOS due to the FM ordering. 
The unique magnetic structure of antiferromagnetically alternating FM CeTe
layers cancels the internal fields on the Te sheets, which become
superconducting. This fact negates the crucial negative effect of FM on SC.
All these factors make CeTe$_{1.82}$ a remarkably high $T_{SC}$ = 2.7 K 
among $f$-electron systems. Finally, the pressure is required to increase 
the interlayer coupling and $T_{SC}$ in the layered superconductors by decreasing 
the interlayer distance along the $c$ axis.

We are grateful for helpful discussions with G. S. Boebinger, A. H. Castro Neto, 
H. Y. Choi, Z. Fisk, S. I. Lee,  A. J. Millis, J. L. Sarrao, J. D. Thompson, 
and C. M. Varma. We are indebted to J. Kamarad (Czech Acd. of Sciences) who made the 
pressure cell at NHMFL-Los Alamos. This work was supported by a grant from NSF 
(DMR-0094241). Work at NHMFL was performed under the auspices of the NSF, 
the State of Florida, and the US Department of Energy.

\begin{center}
\begin{figure}
\epsfig{figure=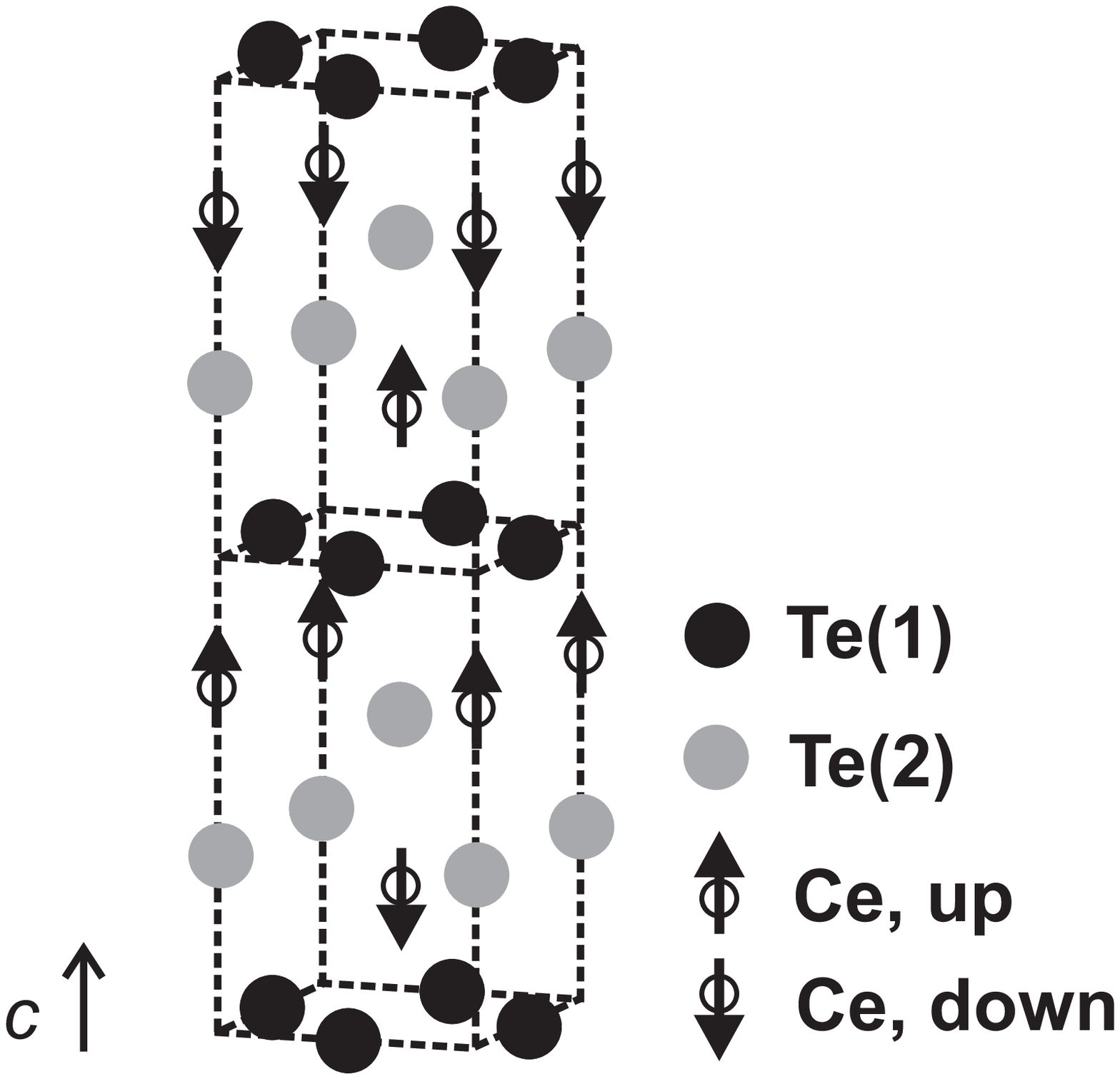, width=0.7\linewidth} 
\caption{The crystal and
magnetic structure of CeTe$_2$ below $T_{N}$. \label{fig1}}
\end{figure}
\end{center}

\begin{figure}
\epsfig{figure=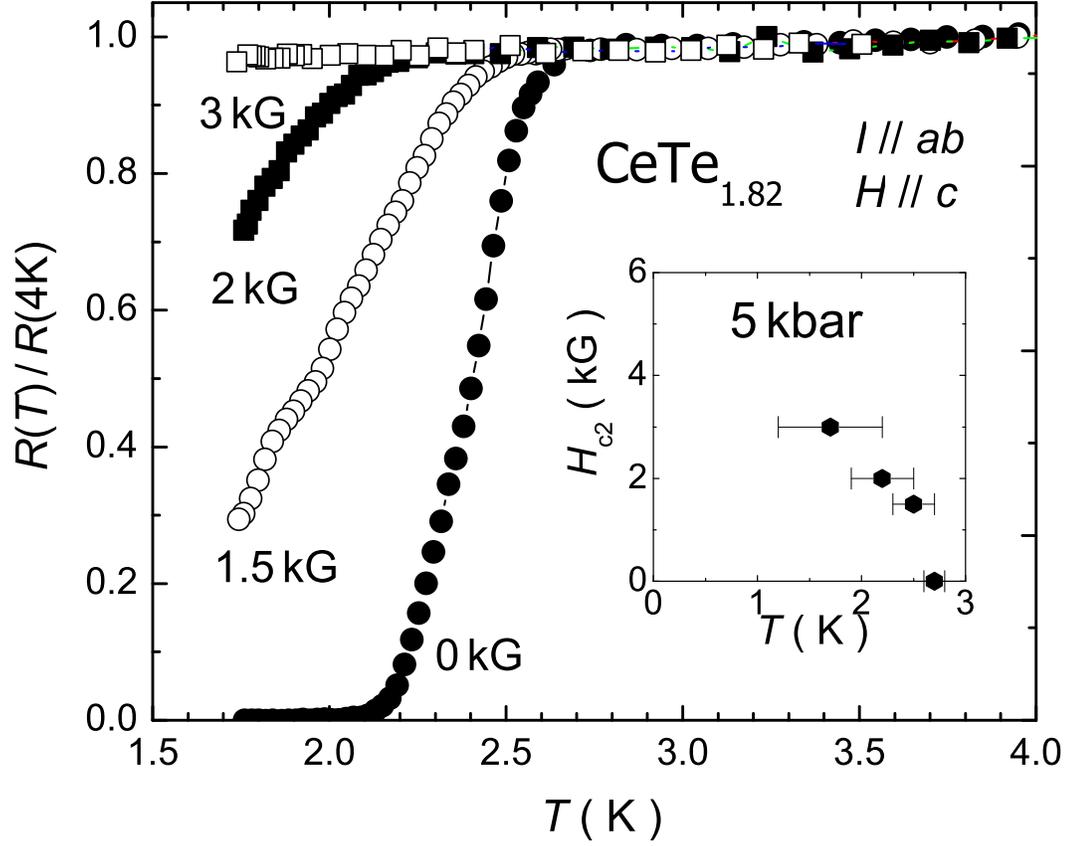,width=1.0\linewidth} 
\caption{Temperature dependence of
the in-plane resistance R(T)/R(4K) normalized to 4 K value in pressure 5 kbar
at c-axis fields 0, 1.5, 2 and 3 kG. The inset shows upper critical fields
$H_{c2}(T)$ from the onset of superconducting transition, at which the
resistance first deviates from the normal-state value. \label{fig2}}
\end{figure}

\begin{figure}
\epsfig{figure=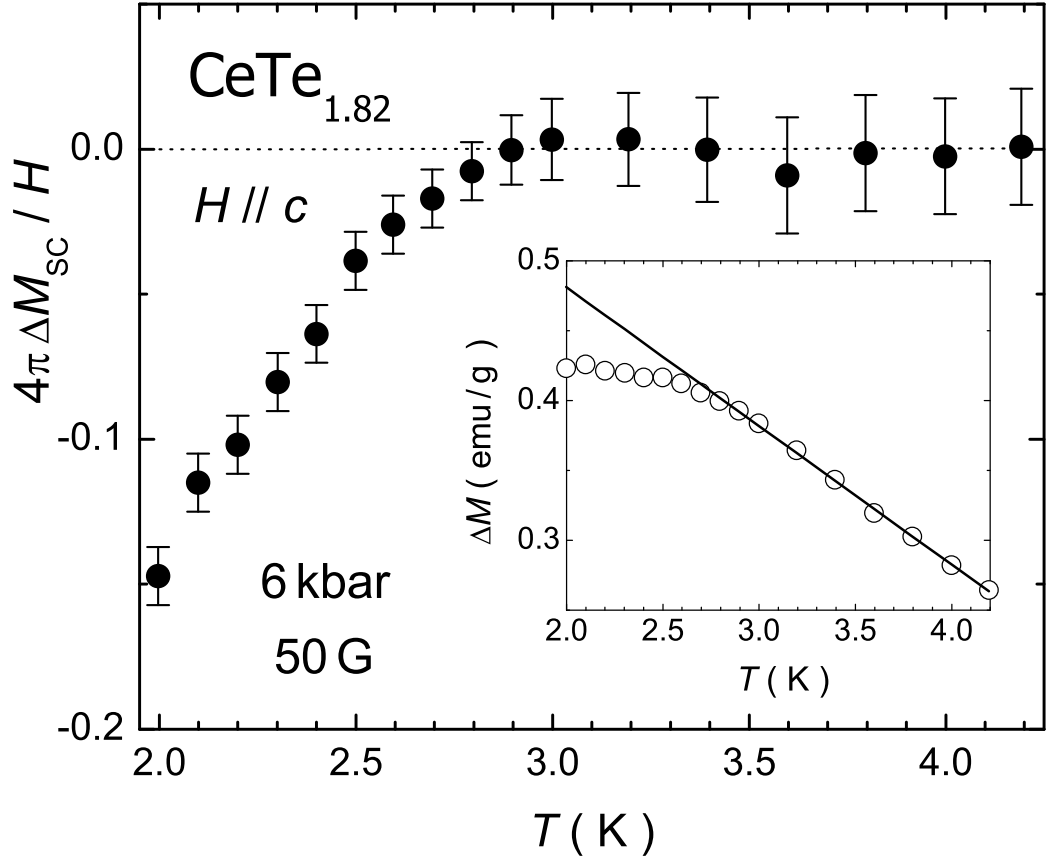,width=1.0\linewidth} 
\caption{Temperature dependence of $4 \pi \Delta M_{SC} / H$ (main panel), 
$\Delta M$ (open cicles in the inset), and $M_{BG}$ (straight line in the inset) at 6 kbar, 
defined in the text: $\Delta M=M_{ZFC}-M_{FC}$ and $\Delta M_{SC}=\Delta M-M_{BG}$.
Density of CeTe$_{1.82}$ under pressure is assumed about 10 g/cc. \label{fig3}}
\end{figure}

\begin{figure}
\epsfig{figure=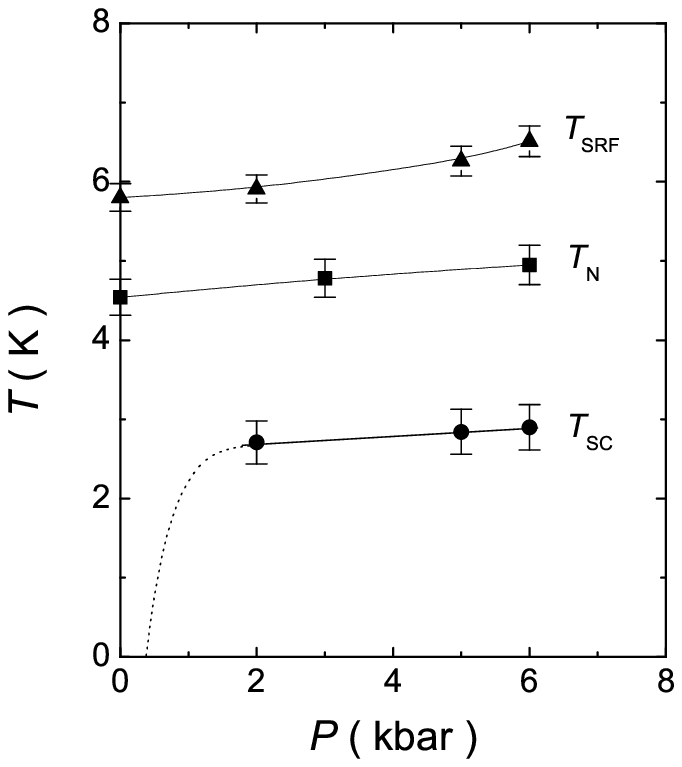,width=1.0\linewidth} 
\caption{Various critical
temperatures as a function of applied pressure:  the short-range ferromagnetic
ordering temperature $T_{SRF}$, the long-range antiferromagnetic ordering
temperature $T_{N}$, and the superconducting transition temperature $T_{SC}$.
$T_{SRF}$ and $T_{SC}$ are determined from $\rho (T, P)$ data and $T_N$
from $M(T, P)$ data. The solid and dotted lines are guides for eyes. \label{fig4}}
\end{figure}



\begin{references}
\bibitem{DOS}
The linear Sommerfeld coefficient $\gamma \sim$ 50 mJ/K$^2$mol estimated by specific heat 
measurement well reflects a lower density of states compared to other heavy fermion 
compounds. Also from the Hall coefficient measurement the carrier density is extremely 
low $\sim 0.001$ per formula unit, although this low value should reflect a large
cancellation between hole-like and electron-like bands.
\end{references}
\end{document}